\documentclass[intlimits,twoside,a4paper]{article}

\usepackage[cp1251]{inputenc}

\usepackage[eqsecnum]{cmpj3}

\usepackage{bm}

\issue{2022}{25}{4}{43705}
\doinumber{10.5488/CMP.25.43705}

\title[Dielectric relaxation induced by oxygen vacancies in Na$_{0.5}$Bi$_{0.5}$TiO$_{3}$ ceramics]%
{Dielectric relaxation induced by oxygen vacancies in Na$_{0.5}$Bi$_{0.5}$TiO$_{3}$ ceramics%
}
\author[V. M. Sidak, M. P. Trubitsyn, T. V. Panchenko]{V. M. Sidak\orcid{0000-0002-2691-1836}\refaddr{label1}\thanks{Corresponding author: \email{vasylsidak@gmail.com}.},
        M. P. Trubitsyn\orcid{0000-0001-7993-7733}\refaddr{label2}, T. V. Panchenko\orcid{0000-0003-2890-2893}\refaddr{label2}}
\addresses{
\addr{label1} Dnipro State Medical University, 9 Vernadsky St., 49044 Dnipro, Ukraine
\addr{label2} Oles Honchar Dnipro National University, 72 Gagarina Ave., 49045 Dnipro, Ukraine 
}

\date{Received June 30, 2022, in final form August 31, 2022}

\begin{document}

\maketitle

\begin{abstract}	
	
Dielectric permittivity was studied in ceramics of relaxor ferroelectric bismuth-sodium titanate Na$_{0.5}$Bi$_{0.5}$TiO$_{3}$. The measurements were performed on as sintered and heat treated in vacuum samples. The diffuse dielectric anomalies associated with the structural phase transitions were observed in as sintered samples. The intense peak of permittivity (\( \varepsilon_{\text{max}} \sim 10^{4}\)) appeared after heat treating in vacuum. The anomaly of \(\varepsilon(T)\) was contributed by slow polarization processes (\(f<10\)~kHz) and was non-stable, vanishing on heating in air up to $\sim$~800 K. Temperature and frequency dependencies of \(\varepsilon\) were described by using Cole-Cole model with accounting thermally stimulated decay of the non-stable polarization. It is supposed that the dielectric anomaly is determined by space charge polarization mechanism. Oxygen vacancies  V$_{\textsc{O}}^{\bullet \bullet}$ and electrons localized on titanium ions Ti$'_{\textsc{Ti}}$ 
 are assumed to be responsible for the phenomenon observed. 
%
%
%\keywords Up to six keywords (\href{https://physh.aps.org/browse}{Physics Subject Headings})
\printkeywords
%
%\pacs 77.22.-d, 77.84.-s, 77.84.Dy
\end{abstract}

\section{Introduction}

%\doclicenseThis

High sensitivity for external fields is the most valuable requirement for functional materials used to transform energy from certain kind to another one. An increased susceptibility often results from lattice instability in the range of structural phase transition. That is why crystalline compounds undergoing structural transformations are intensively investigated by researchers and technologists involved in creation of new functional materials for piezoelectric, thermoelectric, photovoltaic and other converters. The crystals with perovskite ABO$_{3}$ structure have found wide range of applications in modern electronics. Consequently, the compounds of the perovskite family are among the most popular objects for studies in materials sciences. Variations of chemical composition, formation of the structure on nano- and micro\-meter levels, control on the lattice defects make it possible to create the materials with a broad variety of physical properties. Thus, ceramics based on Pb-ZrTiO$_{3}$ show extremely high electro-mechanical para\-meters and are used in piezoelectric devices \cite{Bobic2018}. Introducing the transition groups ions into the structural ABO$_{3}$ unit leads to the appearance of magneto-electrical coupling in multiferroic materials (BiMnO$_{3}$, BiFeO$_{3}$, TbMnO$_{3}$) \cite{Glinchuk2013}. Some crystals with complex perovskite structure like ACu$_{3}$Ti$_{4}$O$_{12}$ ($\text{A}=\text{Ca}$, Ba, Sr) possess extremely high dielectric constants $(\sim10^{4}-10^{5})$ which opens new prospects to be used as the materials with high permittivity in memory and microwave devices \cite{Singh2014,Zhang2021}. 

It is well known that structural imperfections can strongly affect the properties of crystals and even are capable of inducing new phenomena which are not observed in a perfect lattice.  That is why comprehensive information on typical intrinsic and extrinsic lattice defects becomes of high importance. At the present time, numerous works are aimed at studying the mechanisms of the influence of defects  on the properties of crystals. Based on the knowledge gained, the technological approaches are developed that allow to control qualitatively and quantitatively the defectiveness of the crystal structure. Doping with iso- or heterovalent impurities, heat treatment in various atmospheres, applying external fields make it possible to stimulate the appearance of the defects that improve the targeted characteristics or, conversely, to reduce the content of undesirable defects that degrade the useful parameters. Intensive experimental and technological studies aimed at controlling the  subsystem of defects, have led to the appearance of the ``defect engineering'' concept~\cite{Rudolph2016}.

\begin{figure}[!t]
%	\vspace{-2ex}
	\centerline{\includegraphics{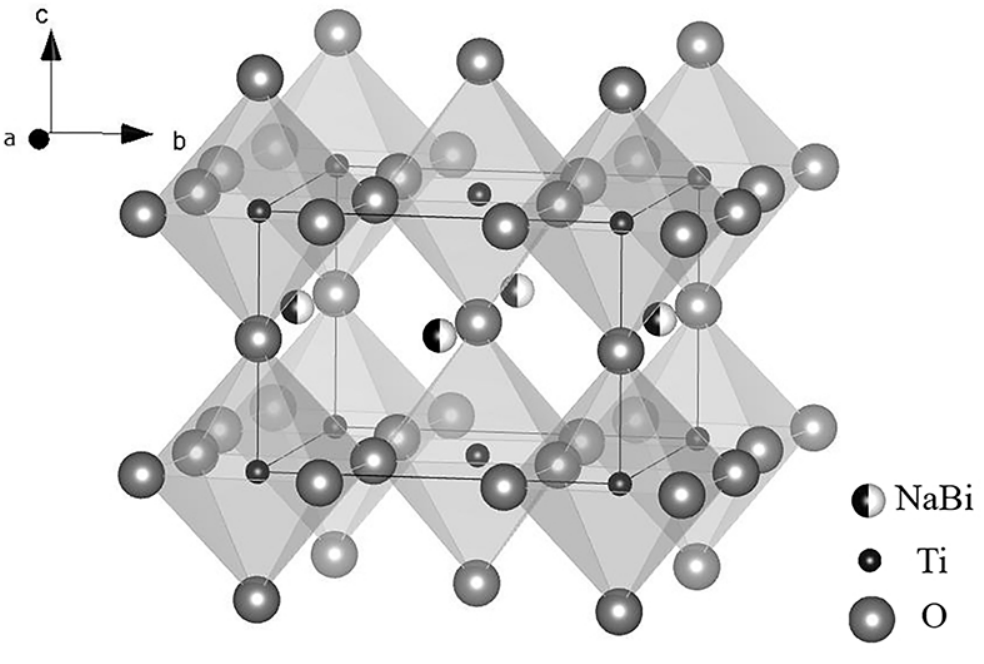}}
	\caption{The crystal structure of NBT in tetragonal phase \cite{Jones2002}.}
	\label{fig-1}
\end{figure}

Modern requirements in the field of environmental protection considerably changed the situation in the production of functional materials and urge the search for new compositions free from health harmful chemical elements. Lead-free bismuth-sodium titanate Na$_{0.5}$Bi$_{0.5}$TiO$_{3}$ (NBT) meets these requirements and shows a number of attractive physical properties. NBT crystal belongs to a group of complex perovskites with the structure of A'A''BO$_{3}$ type, where sodium and bismuth atoms are randomly distributed through the A-site (figure~\ref{fig-1}) \cite{Jones2002}. The extremely high electro-mechanical coupling is the most prominent physical property of NBT crystal and solid solutions based on it \cite{Priya2012}. The specific properties are directly related to the structural phase states observed in NBT. On cooling from high-temperature side NBT undergoes the following sequence of phase transitions: from cubic to tetragonal ferroelastic phase at \(T_{C} \approx 810~\)K, and further to rhombohedral ferroelectric phase at \(T_{R} \approx 490~\)K \cite{Jones2002}. In the range of T$_{R}$, NBT demonstrates high permittivity and specific dielectric dispersion peculiar to relaxor ferroelectrics~\cite{Isupov2003}. Besides, the properties of NBT can be substantially modified  by doping and technological treatments~\cite{Kruzina2014,Kruzina2018,Suchanicz2017,Suchanicz2021}.

Recently, the strong dielectric anomaly (\(\sim 10^{4}\)) was observed near 670--690 K in NBT single crystal~\cite{Kruzina2014,Sidak2015,Sidak2020,Sidak2022} and Na$_{0.5}$Bi$_{0.5}$TiO$_{3}$ – BaTiO$_{3}$ (NBT–BT) solid solutions \cite{Sidakinp}. The \(\varepsilon(T)\) dependence showed an anomalous temperature behaviour and unusual frequency dispersion (here and below symbol~\(\varepsilon\) without prime means real part of permittivity). In addition, permittivity peak disappeared after heat treatment in air ($\sim 800$~K) and could be restored by heat treating in vacuum (\(\sim 1070\) K). The authors of~\cite{Sidak2022} supposed that dielectric anomaly was contributed by the dipole defects formed by oxygen vacancies (V$_{\textsc{O}}^{\bullet \bullet}$) and electrons localized on the nearest titanium ions Ti$'_{\textsc{Ti}}$. The associated dipole defects (Ti$'_{\textsc{Ti}}$-V$_{\textsc{O}}^{\bullet \bullet}$)$^{\bullet}$  were considered as unstable and decomposing upon heating.

These results were obtained for NBT single crystals. Of course, for practical applications, NBT ceramics can be expected as more commercially and technologically acceptable. In this paper anomalous dielectric relaxation mentioned above is studied in NBT ceramics. By accounting the permittivity value in maximum $(\sim10^{4})$, the previous interpretation based on the dipole defects \cite{Sidak2022,Sidakinp} is considered critically. It is supposed that a strong dielectric peak can be associated with space charge polarization phenomenon. The possible microscopic mechanisms of the dielectric relaxation are briefly discussed. 
\newpage

\section{Experimental results}\label{sec2}

The NBT ceramics were prepared by usual sintering technique. The samples for electrical properties measurements were cut off as the plane-parallel plates with the edges of about \(5\times5\times0.8\) mm$^{3}$. The Pt electrodes were deposited on the main planes of the samples by cathode sputtering method. Electrical properties were measured  using AC bridge P 5083 in the temperature interval 300--800 K for the frequency range 0.5--100 kHz. Two types of the samples were used: i) prepared from as sintered ceramics and ii) heat treated in vacuum. The regimes of heat treating were the same as those previously used for single crystals ($T$ = 1070 K, $t$ = 2 h, \(p \approx 1\) Pa) \cite{Sidakinp}. 

\begin{figure}[htb]
	\centerline{\includegraphics[width=0.65\textwidth]{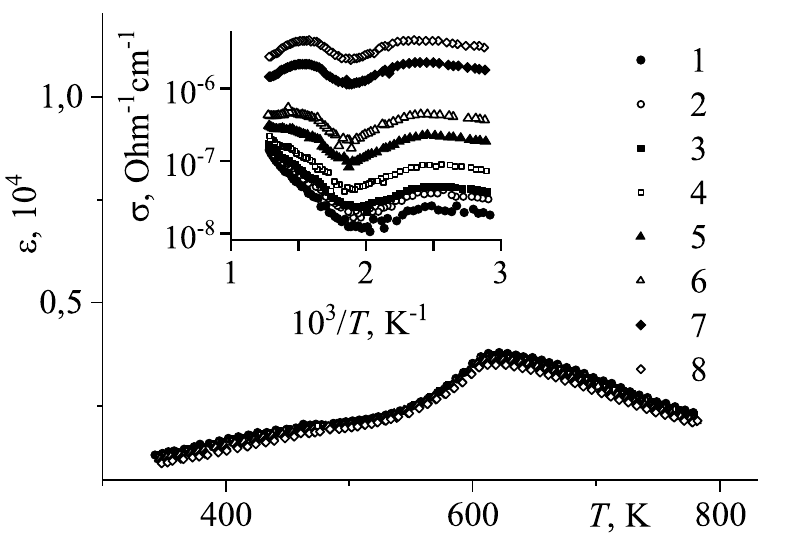}}
	\caption{The permittivity dependencies \(\varepsilon(T)\) in as sintered NBT ceramics. The AC field frequency was f = 0.5 (1); 0.8 (2); 1 (3); 2 (4); 5 (5); 10 (6); 50 (7); 100 (8) kHz. The inset shows Arrhenius plot of conductivity \(\sigma(1/T)\) dependencies.} 
	\label{fig-2}
\end{figure}

The temperature dependencies of dielectric permittivity \(\varepsilon\) and electrical conductivity \(\sigma\) measured on heating for as sintered NBT ceramics are shown in figure \ref{fig-2}. In contrast to the data obtained for NBT and NBT-BT single crystals \cite{Sidak2022,Sidakinp}, \(\varepsilon(T)\) dependence does not show intense relaxation anomaly and reflects the structural transformations in the range of T$_{C}$ and T$_{R}$ only (figure~\ref{fig-2}). The \(\varepsilon(T)\) dependencies measured on the next cooling run and on the subsequent heating-cooling cycles  coincide with each other. The inset to figure \ref{fig-2} shows the temperature dependencies of conductivity $\sigma$ plotted in Arrhenius scale. One can see that $\sigma$ increases with AC field frequency $f$ and weakly depends on temperature. This behaviour is typical of dielectrics at relatively low temperatures. Only at frequencies $f$~<~2~kHz and for \(T~\geqslant 500\)~K conductivity starts to grow exponentially on heating that gives nearly linear regions in the Arrhenius plot. Such a behaviour reflects a growing contribution of thermally activated charge transfer.
\begin{figure}[htb]
	\centerline{\includegraphics[width=0.50\textwidth]{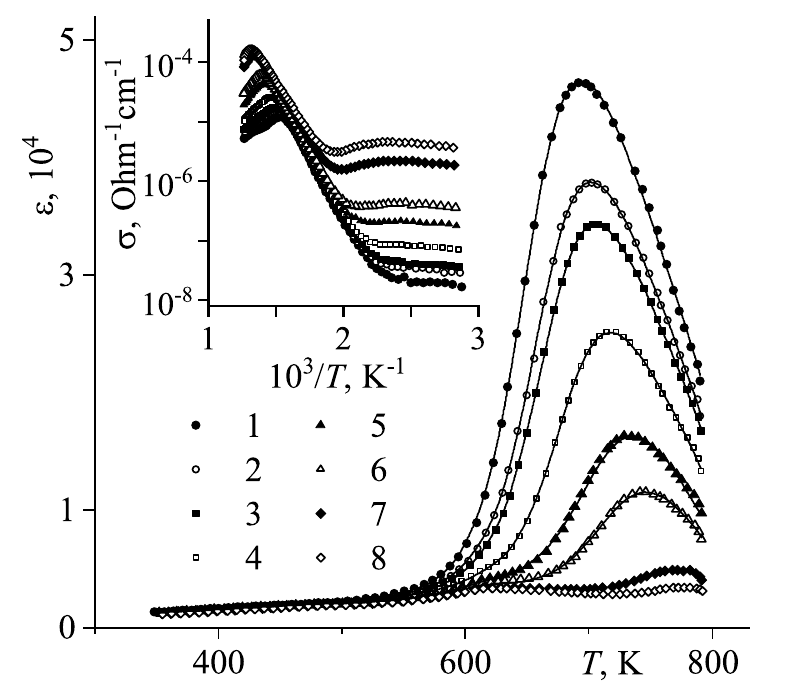}}
	\caption{The dependencies \(\varepsilon(T)\) measured in the first heating run of NBT ceramics previously heat treated in vacuum (1070~K, 2~h). The AC field frequencies are indicated in the caption to figure~\ref{fig-2}. The solid lines were calculated by using (\ref{Coul}), (\ref{Curie2}). The inset shows corresponding \(\sigma(1/T)\) dependencies.} \label{fig-3}
\end{figure}

Next, the sample of NBT ceramics was heat treated in vacuum, cooled to room temperature and after that its electrical properties were measured. The data obtained are shown in figure~\ref{fig-3}. One can see that after heat treating, in the range 700 – 780 K, \(\varepsilon(T)\) demonstrates intense maximum (\(\varepsilon_{\text{max}} \sim 5\cdot10^{4}\), $f = 0.5$~kHz) which is strongly dependent on frequency $f$. As $f$ increases, the peak of  \(\varepsilon(T)\) sharply decreases in magnitude and shifts to higher temperatures. Similarly to the data observed for NBT and NBT-BT single crystals \cite{Sidak2022,Sidakinp}, intense \(\varepsilon(T)\) maximum (figure~\ref{fig-3}) could be detected for the first heating run only and disappeared for the next cooling and heating runs. Corresponding dependencies of conductivity \(\sigma(1/T)\) are shown in the inset to figure~\ref{fig-3}. In the low-temperature interval ($T<500$~K), \(\sigma\) demonstrates nearly the same behaviour as in the untreated sample (the inset to figure~\ref{fig-2}), but for higher temperatures conductivity shows an intense peak corresponding to the relaxation maximum of \(\varepsilon(T)\). In subsequent temperature runs, the \(\sigma(1/T)\) dependencies did not show contribution from dielectric relaxation and were the same as shown in the inset to figure~\ref{fig-2}. 

\section{The model}\label{sec3}

As mentioned in section~\ref{sec2}, the dielectric anomaly \(\varepsilon(T)\) similar to the one shown in figure~\ref{fig-3} was earlier detected in single crystals of NBT and NBT-BT \cite{Sidak2022,Sidakinp}. Special attention was paid to the nearly symmetrical shape of permittivity peak, that was quite different from the asymmetrical \(\varepsilon(T)\) anomaly of Debye relaxator. It was proposed that dipoles or associated complexes responsible for the dielectric anomaly were thermally destroyed on heating. Such decomposition was noticeable in the temperature range where the dielectric relaxation was detected. Consequently, the high-temperature wing of the \(\varepsilon(T)\) anomaly decreased more sharply.

The first attempt to explain the specific character of the dielectric anomaly (figure~\ref{fig-3}) was made in~\cite{Sidak2022} where a decrease of the dipoles or consentration of mobile defects  was described as simple exponential temperature decay. The possible role of configurational and vibrational entropy of the dipole defects was considered somewhat later in \cite{Sidak2020}. Nevertheless, these approaches allowed to interpret the data only at the qualitative level and not provide a correct quantitative description of the experimental results. More accurately, the \(\varepsilon(T, f)\) behavior in NBT-BT single crystal was described in \cite{Sidakinp}, where Debye relaxator model was combined with the kinetic equation that determins the decay of the polarizing entities with temperature.

%Really, in an external AC field permittivity behavior can be described by Cole-Cole formulae \cite{Poplavko2020}
Dielectric response of real structures in an external AC field can be described by Cole-Cole, Davidson-Cole and other models \cite{Poplavko2020}. These models predict different types of dielectric spectra, symmetrical or non-symmetrical diagrams in complex ($\varepsilon{'}$--$\varepsilon{''}$) plane, where $\varepsilon{'}$ and $\varepsilon{''}$ represent real and imaginary parts of permittivity. The dielectric anomaly shown in figure~\ref{fig-3} is observed practically in the same temperature-frequency range ($T>500$~K, $f<10$~kHz), where charge transfer processes notably contribute to conductivity (the nearly linear regions in the \(\sigma(1/T)\) dependencies, the insert to figure~\ref{fig-2}). That is why the experimental diagrams plotted in ($\varepsilon{'}$--$\varepsilon{''}$) for the used frequency region do not permit to make a reliable choice between the models mentioned. Hence, the experimental data are described by Cole-Cole model \cite{Poplavko2020}

\begin{align}
\label{Coul}
\varepsilon^{\ast}(T, \omega) = \varepsilon_{\infty}+\frac{C/T}{1+(\ri \omega \tau_{R})^{1-\alpha}},
\end{align}
where \( \omega = 2\piup f \) is an AC field frequency; $ k $ is Boltzmann constant. Expression (\ref{Coul}) includes a minimum number of the fitting parameters: $\varepsilon_{\infty}$ – permittivity at high-frequency; Curie constant $C\sim n$ which is directly proportional to the concentration $n$ of the dipoles; \( \tau_{R}(T) = \tau^{0}_{R}\exp(E/kT) \) is the time of the relaxation of dipole moments  in an external field; energy parameter $E$ estimates the height of the potential barrier which is overcome at the reorientation dipole moments; phenomenological parameter $0 \leqslant \alpha < 1$  describes the distribution of relaxation times $\tau_{R}$ in disordered structures. 

It should be noted that for the samples heat treated in vacuum (figure~\ref{fig-3}), the anomalies of permittivity imaginary part $\varepsilon{''}(T,~f)$ contain contributions from dielectric relaxation and charge transfer in the same temperature-frequency range (see the comments above). Hence, the analysis of $\varepsilon{''}(T,~f)$ dependences needs to separate these contributions with apriori unknown parameters. The anomalies of permittivity real part $\varepsilon{'}(T,~f)$ (figure~\ref{fig-3}) are mainly contributed by the polarization processes. Thus, the parameters of the discussed dielectric relaxation can be determined more directly from $\varepsilon{'}(T,~f)$ dependencies. In addition, non-stable nature of polarization causes a specific type of anomalous behaviour which is more evident just for dependencies $\varepsilon{'}(T)$. That is why the following analysis is focused on permittivity real part dependencies. 

Further, it should be considered that polarizing entities (dipoles, associated complexes) are non-equilibrium and undergo thermal decomposition. One can assume that a decrease of their concentration $n$ can be described by the simple kinetic equation \cite{Chen1981}
\begin{align}
\label{kinetic}
\frac{\rd n}{\rd t} = - \frac{n}{\tau_{D}} .
\end{align}
Here, \(\tau_{D}(T) = \tau^{0}_{D}\exp(U/kT) \)  and $U$ are the time and energy parameters determining the thermal decay of non-stable polarization. In (\ref{kinetic}), one can go from differentiation in time to a derivative in temperature by considering that during the experiments, the samples were heated and cooled with a constant rate. Thus, the temperature of the samples can be written as \( T(t) = T_{0}+\gamma t \) , where T$ _{0}$ is an initial temperature; $\gamma$ is the rate of temperature changes; $t$ is the current time. Thus, remembering that $C \sim n$, from equation~\ref{kinetic} one can rewrite Curie constant as 
\begin{align}
\label{Curie1}
C(T) = C_{0} \cdot \exp\left[-\frac{1}{\gamma \tau^{0}_{D}} \cdot \int_{T_{0}}^{T} \exp\left(-\frac{U}{kT}\right) \rd T\right] .
\end{align}
%
%$C(T)$ can be calculated from \ref{Curie1} numerically or alternatively the approximate solution of \ref{Curie1} \cite{Kazukauskas2005} can be used
Direct fitting of expression (\ref{Coul}), with accounting (\ref{Curie1}), to the experimental data was complicated and did not give reliable results. Nevertheless, an approximate integration in (\ref{Curie1}) performed in \cite{Simmons} yielded the following expression

\begin{align}
\label{Curie2}
C(T) = C_{0} \cdot \exp\left[-\frac{kT^{2}}{\gamma \tau^{0}_{D}(U+kT)} \exp\left(-\frac{U}{kT}\right) \right].
\end{align}

Thus, the experimental data shown in figure~\ref{fig-3} can be described  using Cole-Cole formulae \ref{Coul} combined with the approximate solution \ref{Curie2} of the kinetic equation \ref{kinetic}.

\section{Discussion}

It is well known that oxygen vacancies V$_{\textsc{O}}^{\bullet \bullet}$ are the typical defect for the crystals of complex oxides. In tightly packed structures like perovskites ABO$_{3}$, the excess positive charge (+2\textit{e}) associated with V$_{\textsc{O}}^{\bullet \bullet}$ more probably can be compensated by the necessary number of cationic vacancies, which gives rise to the appearance of Schottky-type defects. If the concentration of V$_{\textsc{O}}^{\bullet \bullet}$ is too high with respect to the number of cation vacancies, the additional electronic defects appear and the valence of the cations neighboring the V$_{\textsc{O}}^{\bullet \bullet}$ can decrease. In ABO$_{3}$ structures, the weakly bound electrons can be localized on titanium ions and as a result, Ti$'_{\textsc{Ti}}$ centers can arise \cite{Erdem2010,Yang2018,Scharfschwerdt1996}. The energy levels of Ti$'_{\textsc{Ti}}$ centers are shallow enough, and the electrons that hop  via regular titanium ions can participate in the charge transfer. The presence of the nearest neighboring V$_{\textsc{O}}^{\bullet \bullet}$ stabilizes the localized electrons and as a result, associated pairs (Ti$'_{\textsc{Ti}}$-V$_{\textsc{O}}^{\bullet \bullet}$)$^{\bullet}$ can arise \cite{Erdem2010}.

One can expect that thermal treatment of NBT ceramics in vacuum (\textit{T} = 1070 K) should increase mainly the concentration of V$_{\textsc{O}}^{\bullet \bullet}$. Each  V$_{\textsc{O}}^{\bullet \bullet}$ that arose in the treated ceramics could cause the emergence of two Ti$'_{\textsc{Ti}}$ centers. Correspondingly, the appearance of  intense \(\varepsilon(T)\) anomaly after heat treatment (figure~\ref{fig-3}) can be  just associated with the defects formed by V$_{\textsc{O}}^{\bullet \bullet}$. That is why in the previous works \cite{Kruzina2014,Sidak2020,Sidak2022,Sidakinp} a slow dielectric relaxation in NBT single crystals was attributed to re-orientations of (Ti$'_{\textsc{Ti}}$-V$_{\textsc{O}}^{\bullet \bullet}$)$^{\bullet}$ dipoles resulting from hopping of V$_{\textsc{O}}^{\bullet \bullet}$ through oxygen octahedra vertices. Thermal decay of polarization (\ref{kinetic}) was interpreted as a result of disassociation of (Ti$'_{\textsc{Ti}}$-V$_{\textsc{O}}^{\bullet \bullet}$)$^{\bullet}$ centers  occurring on heating. Nevertheless, a great value of permittivity in the peak ($ \sim 5\cdot 10^{4}$ at $f=0.5$~kHz, see figure~\ref{fig-3}) can be hardly attributed to the dipole defects, the concentration of which is assumed to be low enough. More probably, such a  high value of \(\varepsilon\) can be the result of space charge polarization which is often observed in inhomogeneous media. Usually, permittivity of such substances is defined as an effective one. 

Let us consider  more in detail the assumption that intense \(\varepsilon(T)\) peak in figure~\ref{fig-3} is contributed by mobile charge defects which in an external electric field can accumulate near certain inhomogeneities.

Earlier in \cite{Kruzina2014,Sidak2020,Sidak2022,Sidakinp} we supposed that reorientation of the dipole complexes (Ti$'_{\textsc{Ti}}$-V$_{\textsc{O}}^{\bullet \bullet}$)$^{\bullet}$ in an external field and their thermal dissociation on heating occurred through V$_{\textsc{O}}^{\bullet \bullet}$ hopping. That is why calculating the \(\varepsilon(T, f)\) dependencies by using (\ref{Coul}),  (\ref{Curie2}), we associated the pre-exponential factors $\tau^{0}_{R}$, $\tau^{0}_{D}$ for the relaxation times with inverse Debye frequency. Correspondingly, the values of $\tau^{0}_{R}$, $\tau^{0}_{D}$ were fixed ($\approx2 \cdot 10^{-13}$~s) and estimated from Debye temperatures ($\theta \approx$~260–350~K) typical of perovskites \cite{Shebanovs2002}.
\begin{figure}[htb]
	\centerline{\includegraphics[width=0.50\textwidth]{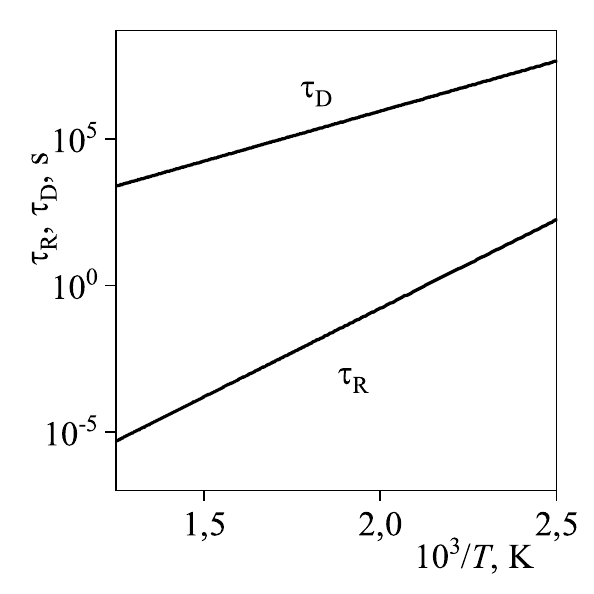}}
	\caption{The dependencies of the dielectric relaxation time $\tau_{R}(1/T)$ and polarization decay time $\tau_{D}(1/T)$ calculated from the data given in table~\ref{tbl-1}.} \label{fig-4}
\end{figure}

Assuming the space charge polarization to be the main effect, we have no apriori information on the values of $\tau^{0}_{R}$, $\tau^{0}_{D}$  and  that is why we set them free. We should add that considering $\tau^{0}_{R}$, $\tau^{0}_{D}$ as the fitting parameters allowed to reduce by about an order of magnitude the mean square deviation of the calculated data from the experimental ones. The curves calculated with the help of the model discussed in section~\ref{sec3}, are drawn in figure~\ref{fig-3} with the solid lines. It should be noted that the background contribution to \(\varepsilon(T)\) due to the structural phase transitions was taken into account as it was described earlier in \cite{Kruzina2014,Sidak2020,Sidak2022,Sidakinp}. In the scale chosen, this contribution shows only a weak dependency on temperature and in pure form can be seen in figure~\ref{fig-2} where intense \(\varepsilon(T)\) peak is absent. The values of the parameters, used in (\ref{Coul}), (\ref{Curie2}) and averaged for all studied frequencies, are presented in table~\ref{tbl-1}. One can see that fitting the calculated data to the experimental ones, in contrast to the previous assumption in \cite{Sidakinp}, gives strongly different values for the pre-exponential factors $\tau^{0}_{R}$ and $\tau^{0}_{D}$. The value of $\tau^{0}_{R}$ corresponds to the order of typical lattice frequencies. By contrast, the factor $\tau^{0}_{D}$ for polarization decay is found to be twelve orders longer which corresponds to the infra-low frequency range. Seemingly, such extremely high value of $\tau^{0}_{D}$ indirectly evidence in favor of space charge polarization mechanism. 

\begin{table}[htb]
\caption{The values of the parameters in (\ref{Coul}), (\ref{Curie2}) obtained from the \(\varepsilon(T,~f)\) dependencies (figure~\ref{fig-3}).}
	\label{tbl-1}
	\vspace{1ex}
	\begin{center}
		\renewcommand{\arraystretch}{0}
		\begin{tabular}{|c|c|c|c|c|c|}
			\hline\strut
			C$_{0}$,~K&$\alpha$&$\tau^{0}_{R}$,~s&$E$,~eV&$\tau^{0}_{D}$,~s&$U$, eV\\
			\hline
			\rule{0pt}{2pt}&&&&&\\
			\hline
			$3.4(8)\cdot10^{7}$& $0.09(1)$& $1.2(2)\cdot10^{-13}$& 1.21(1)& $1.1(2)\cdot10^{-1}$&  0.67(2)\strut\\
			\hline			
		\end{tabular}
		\renewcommand{\arraystretch}{1}
	\end{center}
\end{table}

The dependencies of dielectric relaxation time $\tau_{R}(1/T)$ and polarization decay time $\tau_{D}(1/T)$, calculated with the help of the data in table \ref{tbl-1}, are shown in figure \ref{fig-4}. One can see that for the whole studied interval, $\tau_{D}$ values considerably exceed the typical time ($\sim 1$~s) of a single measurement at certain $T$ and~$f$. On the other hand, $\tau_{D}$ values are comparable with the time ($\sim3-4$~h) of a single measuring run.

\begin{figure}[!t]
	\centerline{\includegraphics[width=0.50\textwidth]{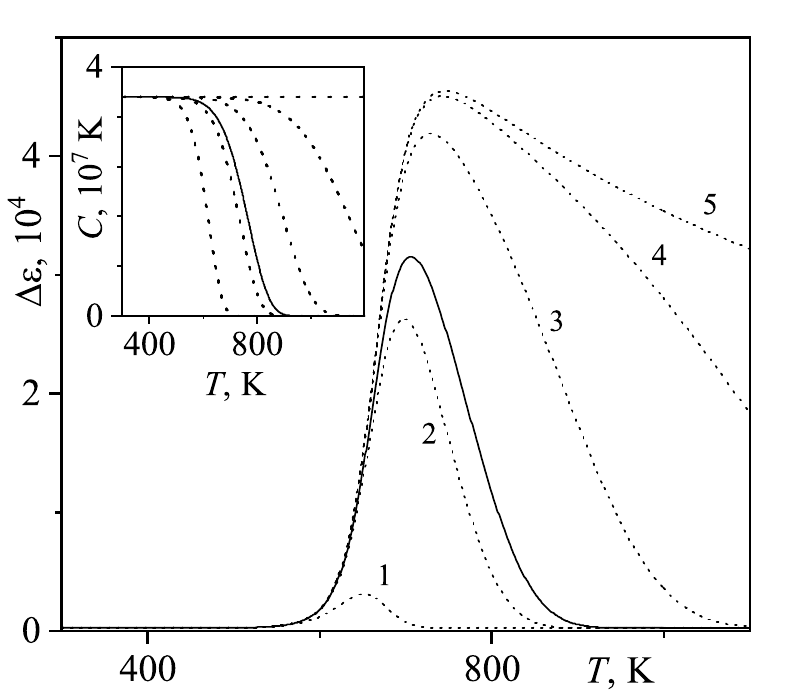}}
	\caption{The contribution from non-stable polarization to dielectric anomaly \(\Delta\varepsilon(T)\). The dashed curves are calculated by using the following heating rates $\gamma$ = 0.1~(1); 1~(2); 10~(3); 10$^{2}$~(4); 10$^{6}$~(5) K/min. The parameters in (\ref{Coul}), (\ref{Curie2}) are taken from processing the experimental data measured at $f = 1$ kHz. The solid line is calculated for the experimental data in figure~\ref{fig-3} ($\gamma$~=~1.7 K/min, $f~=~1$ kHz). The inset shows the corresponding dependencies of Curie constant $C(T)$.} 
	\label{fig-5}
\end{figure}

One can show that the model (section 3) combining Cole-Cole formulae (\ref{Coul}) with kinetic equation~(\ref{kinetic}) allows one to describe the specific features of the dielectric relaxation discussed. Thus, (\ref{Curie2}) predicts that the form of the dielectric anomaly \(\varepsilon(T, f)\) should depend on time and on heating rate $\gamma$. Obviously,  such effects can be tested in experiment. Besides, one can expect that the behaviour of \(\varepsilon(T, f)\) should depend on the ratio between the rates of dielectric relaxation $\tau^{-1}_{R}$  and polarization decay $\tau^{-1}_{D}$. Really, the anomaly \(\varepsilon(T, f)\) takes a specific form (figure~\ref{fig-3}) since the decay of non-equilibrium polarization becomes notable in the same temperature range where dielectric peak is detected. Thus, one can expect that the type of dielectric anomaly for certain $\tau^{0}_{R}$, $\tau^{0}_{D}$ values should depend on the ratio between activation energies $U/E$. Let us consider the effects mentioned.

\begin{figure}[htb]
	\centerline{\includegraphics[width=0.50\textwidth]{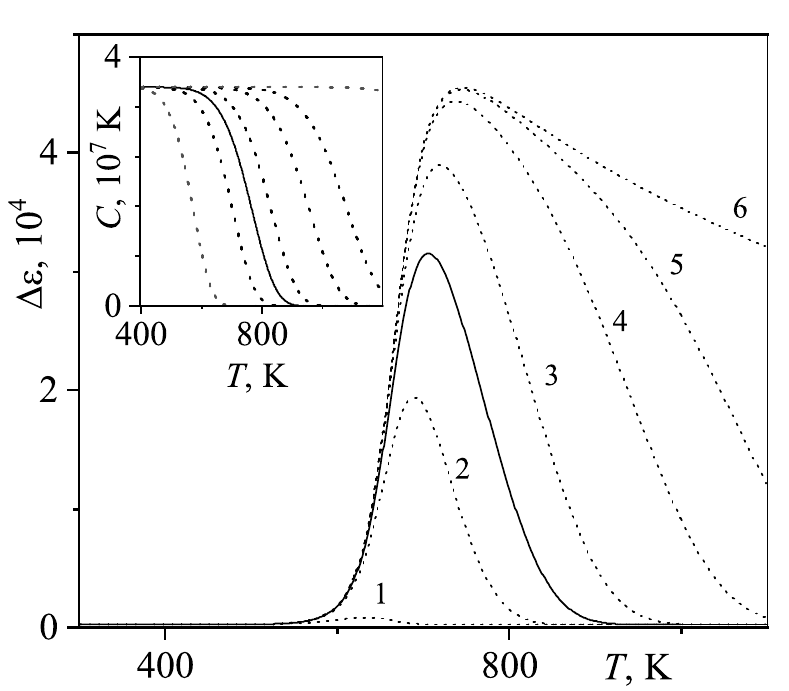}}
	\caption{The dielectric anomaly $\Delta\varepsilon(T)$ (dashed lines) calculated for the following ratios $U/E$ = 0.4~(1); 0.5~(2); 0.6~(3); 0.7~(4); 0.8~(5); 1.2~(6). The value of $E = 1.21$~eV is fixed, and the value of $U$ is varied. The solid line corresponds to the experimental data in figure 3 ($U/E$ = 0.55, $f=1$~kHz). The $C(T)$ dependencies are shown in the inset.} 
	\label{fig-6}
\end{figure}

Figure~\ref{fig-5} shows the \(\varepsilon(T)\) anomaly calculated for different $\gamma$ values. One can see how considerably the variations of $\gamma$ can change the \(\varepsilon(T)\) behaviour. For high values of $\gamma$ during the whole measuring cycle, the decay of non-equilibrium polarization remains practically negligible. Correspondingly, in the limit $\gamma \to \infty$, the behaviour of \(\varepsilon(T)\) approaches a classic behavior of Debye relaxator (figure~\ref{fig-5}). For the intermediate values of $\gamma$, the anomaly of \(\varepsilon(T)\) takes the form of nearly symmetrical peak. At infinitely low heating rate ($\gamma \to 0$), non-equilibrium polarization has enough time to decay totally before the permittivity peak can be detected. As a result, on lowering $\gamma$, the permittivity peak decreases in amplitude and finally disappears (figure~\ref{fig-5}). The inset to figure~\ref{fig-5} shows the calculated temperature dependence of Curie constant. On heating, $C(T)$ shows a step-like decrease manifesting a decay of non-equilibrium polarization. For high rates of~$\gamma$, Curie constant possesses a maximum possible value and is practically temperature independent  in the whole  interval studied. For low values of $\gamma$, Curie constant on heating decreases to zero before the dielectric relaxation can be detected. 

Figure~\ref{fig-6} illustrates how \(\varepsilon(T)\) anomaly changes its form when the ratio between activation energies $U/E$ varies. For higher values of $U/E$, one has a Debye-type behavior of \(\varepsilon\). On lowering the ratio $U/E$, permittivity \(\varepsilon(T)\) takes the intermediate peak-like form and finally it disappears when the ratio $U/E$ decreases. The inset to figure~\ref{fig-6} shows the corresponding dependencies of Curie constant $C(T)$.

Assuming that an intense \(\varepsilon(T)\) peak (figure~\ref{fig-3}) is determined by space charge polarization effects, and for NBT ceramics one can consider the same typical defects such as oxygen vacancies V$_{\textsc{O}}^{\bullet \bullet}$, electrons localized on titanium Ti$'_{\textsc{Ti}}$ and probably associated complexes based on them. V$_{\textsc{O}}^{\bullet \bullet}$ can be assumed to be more heavy defects, whereas localized electrons Ti$'_{\textsc{Ti}}$ can be supposed to be more light ones. The mobile charge defects can accumulate near the following inhomogeneities in NBT ceramics: i) intergrain boundaries; ii) ferroelectric or ferroelastic domains boundaries; iii) near-electrode regions. Presumably, one can expect that during long enough period the oxygen vacancies can accumulate near certain inhomogeneities and form the regions with higher V$_{\textsc{O}}^{\bullet \bullet}$ concentration. In the applied electric field, electrons Ti$'_{\textsc{Ti}}$ move between the regions with increased V$_{\textsc{O}}^{\bullet \bullet}$ concentration. On heating, due to diffusion, the regions with high V$_{\textsc{O}}^{\bullet \bullet}$ concentration dissolve. More information on the nature of the inhomogeneities which can cause space charge polarization in NBT can be obtained from comparison of the experimental data measured for ceramic and single crystalline NBT. This work is in progress at the moment. 

\section{Summary}
 
An intense low frequency anomaly of dielectric permittivity  appeared in NBT ceramics after heat treating in vacuum ($T=1070$~K, $t=2$~h). The corresponding polarization was found to be non-stable and disappeared after heating in air up to $\sim$~800 K. The results of the thermal treatment evidenced that the observed dielectric relaxation was contributed by the defects including oxygen vacancies. Dielectric maxima were detected in the same temperature-frequency range where charge transfer processes gave notable contribution to conductivity in AC field. That is why we could not examine reliably the type of the experimental diagrams plotted in the complex plane of permittivity. The temperature and frequency dependencies of $\varepsilon$ were described on the basis of Cole-Cole model which could be used to describe dielectric relaxation in partially disordered structures. Thermal decay of the non-equilibrium polarization was described using the simple kinetic equation. The analysis was focused on the behaviour of permittivity real part which was mainly contributed by the polarization processes. Combination of Cole-Cole model with kinetic equation allowed us to describe the experimental data with a good accuracy and to predict the evolution of dielectric anomaly under variations of the experimental conditions and characteristics of the phenomena observed. The great value of permittivity in the maximum (\(\varepsilon_{\text{max}} \sim 5\cdot10^{4}\), $f = 0.5$~kHz), observed for NBT ceramics, made doubtful the assumption that the dielectric anomaly could be due to the dipole defects the concentration of which was assumed to be not extremely high. That is why it was supposed that the observed dielectric relaxation was determined by space charge polarization mechanism. Oxygen vacancies V$_{\textsc{O}}^{\bullet \bullet}$ and electrons localized on titanium ions Ti$'_{\textsc{Ti}}$ were assumed to be responsible for the phenomena studied. One can hope that more details on the microscopic mechanism of the thermally non-stable dielectric relaxation can be derived from comparative studies of the electrical properties of NBT single crystals and ceramics treated in atmospheres enriched and depleted in oxygen.

%% Type in your references using {thebibliography} environment 
%% or create them from your bibtex database using cmpj.bst style.

\section*{Acknowledgements}

The study was funded by Ministry of Education and Science of Ukraine according to the research projects No. 0119U100694, No. 0120U102239 and No. 0122U001228.

%\bibliographystyle{cmpj}

%
%% If you have problems with typesetting in ukrainian uncomment lines below.
%
%  \lastpage
%  \end{document}
%\newpage
\ukrainianpart

\title{Діелектрична релаксація, індукована кисневими вакансіями в кераміці Na$_{0.5}$Bi$_{0.5}$TiO$_{3}$}
\author{В. М. Сідак\refaddr{label1}, М. П. Трубіцин\refaddr{label2}, Т. В. Панченко\refaddr{label2}}
\addresses{
\addr{label1} Дніпровський державний медичний університет, Україна, 49044 Дніпро, вул. Вернадського, 9
\addr{label2} Дніпровський національний університет імені Олеся Гончара, Україна, 49045 Дніпро, пр. Гагаріна, 72
}
%
%% якщо автор є один або автори є з однієї установи:
%
%  \author{1й Автор, 2й Автор, \ldots}
%  \address{Інститут\ldots}
%
%%

\makeukrtitle

\begin{abstract}
\tolerance=3000%
Досліджено діелектричну проникність кераміки релаксорного сегнетоелектрика натрій-вісмутового титанату Na$_{0.5}$Bi$_{0.5}$TiO$_{3}$. Вимірювання проводили на необроблених і термічно оброблених у вакуумі зразках. В необроблених зразках спостерігалися розмиті діелектричні аномалії, пов'язані зі структурними фазовими переходами. Інтенсивний пік діелектричної проникності (\( \varepsilon_{\text{max}} \sim 10^{4}\)) з'явився після термічної обробки у вакуумі. Аномалія \(\varepsilon(T)\) була спричинена повільними процесами поляризації (\(f<10\)~кГц) і була нестабільною, зникаючи при нагріванні в повітрі до~800 K. Температурні та частотні залежності \(\varepsilon(T)\) описано за допомогою моделі Коула-Коула з урахуванням термостимульованого затухання нестабільної поляризації. Передбачається, що діелектрична аномалія визначається механізмом поляризації просторового заряду. Кисневі вакансії V$_{\textsc{O}}^{\bullet \bullet}$ та електрони, локалізовані на іонах титану Ti$'_{\textsc{Ti}}$, вважаються відповідальними за спостережуване явище.
\keywords діелектричнi властивості, діелектрична проникність, перовскіти, дефекти

\end{abstract}

\end{document}